\documentclass[aps,footinbib, notitlepage,superscriptaddress, longbibliography,eqsecnum]{revtex4-2}

\usepackage[english]{babel}
\usepackage[utf8]{inputenc}
\usepackage{amsmath}
\usepackage{amsthm}
\usepackage{braket}
\usepackage{amssymb}
\usepackage{graphicx}
\usepackage{hyperref}
\usepackage[normalsize]{subfigure}
\usepackage{dsfont}
\usepackage{xcolor}
\usepackage{geometry}
\usepackage{mathtools}
\usepackage{bbold}
\usepackage{bbm}
\usepackage[capitalize]{cleveref}
\usepackage{changepage}

\DeclareFontFamily{OT1}{pzc}{}
\DeclareFontShape{OT1}{pzc}{m}{it}{<-> s * [1.10] pzcmi7t}{}
\DeclareMathAlphabet{\mathpzc}{OT1}{pzc}{m}{it}

\definecolor{deepchampagne}{rgb}{0.98, 0.84, 0.65}

\providecommand{\st}[1]{_{\text{#1}}}
\providecommand{\sfrac}[2]{#1/#2}

\providecommand{\ut}[1]{^{\text{#1}}}

\def\onehalf{\frac{1}{2}}

\def\bra{\ensuremath{\langle}}
\def\ket{\ensuremath{\rangle}}

\def\tr{\mathrm{tr}}

\def\Imat{\mathbbm{1}}
\def\im{\mathrm{i}}
\def\Av{\bv{A}}

\def\fv{\bv{f}}

\def\yv{\bv{y}}
\def\zv{\bv{z}}

\def\vv{\bv{v}}

\def\sv{\bv{s}}
\def\xv{\bv{x}}
\def\wv{\bv{w}}
\def\Wv{\bv{W}}
\def\Rv{\bv{R}}

\def\Yv{\bv{Y}}
\def\b0{\bv{0}}

\def\Acal{\mathcal{A}}

\def\Hcal{\mathcal{H}}

\def\Lcal{\mathcal{L}}
\def\Herm{\mathcal{M}}

\def\Ocal{\mathcal{O}}

\def\reals{\mathbb{R}}

\def\numM{m}		

\newcommand{\beq}{\begin{equation}}
\newcommand{\eeq}{\end{equation}}
\newcommand{\beqn}{\begin{equation*}}
\newcommand{\eeqn}{\end{equation*}}
\newcommand{\bv}[1]{\mathbf{#1}}

\begin{document}

\title{Kernel-based optimization of measurement operators for quantum reservoir computers}

\author{Markus Gross}
\email{markus.gross@dlr.de}
\author{Hans-Martin Rieser}

\affiliation{Institute for AI Safety and Security, German Aerospace Center (DLR),\\ Sankt Augustin and Ulm, Germany}

\date{\today}

\begin{abstract} 
Finding optimal measurement operators is crucial for the performance of quantum reservoir computers (QRCs), since they employ a fixed quantum feature map. We formulate the training of both stateless (quantum extreme learning machines, QELMs) and stateful (memory dependent) QRCs in the framework of kernel ridge regression. We thus extend the kernel viewpoint of supervised quantum models to recurrent QRCs by deriving an exact Hilbert--Schmidt kernel representation of the optimal readout observable on history space. This approach renders an optimal measurement operator that minimizes prediction error for a given reservoir and training dataset. For large qubit numbers, this method is more efficient than the conventional training of QRCs. We discuss efficiency and practical implementation strategies, including Pauli basis decomposition and operator diagonalization, to adapt the optimal observable to hardware constraints. To demonstrate the effectiveness of this approach, we present numerical experiments on image classification and time series prediction tasks, including chaotic and strongly non-Markovian systems. The developed method can also be applied to other quantum machine learning models.
\end{abstract}

\maketitle

\section{Introduction}

Since quantum systems are inherently disturbed by measurements, finding efficient and optimal ways to extract information is a crucial part of any quantum machine learning (QML) model, and numerous studies have addressed this problem \citep{bae_quantum_2015,huang_predicting_2020,garcia-perez_learning_2021,glos_adaptive_2022,leporini_efficient_2022,gyurik_structural_2023,nakaji_measurement_2023,lee_variational_2024,chen_learning_2025,chen_learning_2025a,liu_you_2025,li_enhancing_2025,lin_adaptive_2025,vetrano_state_2025}.
For variational QML models, measurement operators are often chosen based on hardware constraints or heuristics. Due to the training of the internal layers, this usually does not pose a severe problem for the accuracy of the model. 
However, in the case of QRCs, training happens only at the readout and thus depends crucially on the information extracted by the measurement operator. Finding optimal measurement operators is thus a key challenge in QRC design \citep{ghosh_quantum_2019,mujal_time_2023,innocenti_potential_2023,xiong_fundamental_2023,suprano_experimental_2024}.

In \cite{schuld_supervised_2021}, the notion of an optimal measurement operator has been introduced based on the equivalence between supervised quantum ML models and quantum kernel methods \citep{schuld_quantum_2019,havlicek_supervised_2019,schuld_supervised_2021,huang_power_2021}.
Such optimal measurement operators have subsequently been identified in a number of quantum ML studies \citep{gan_unified_2023,polson_quantum_2023,rodriguez-grasa_neural_2025}, including a recent discussion of double descent in QELM-like models, which employ a fixed quantum feature map \citep{kempkes_double_2026}.
The measurement problem for QELMs from a more general quantum-information perspective has been addressed in \cite{innocenti_potential_2023}.
On the side of classical reservoir computing, various methods have been proposed to construct equivalent kernel representations \citep{hermans_recurrent_2012,dong_reservoir_2020,gonon_reservoir_2022,grigoryeva_infinitedimensional_2025}.
Besides this, no systematic study of optimal measurement operators for QRCs has been performed.

In the present work, we discuss the problem of determining an optimal measurement operator for QRCs using kernel-based methods.
QRCs have been introduced as a hardware efficient and noise-resilient quantum ML approach \citep{fujii_harnessing_2017,fujii_quantum_2020,mujal_opportunities_2021,domingo_taking_2023}, and have been successfully applied in tasks such as image classification \citep{sakurai_quantum_2022,kornjaca_largescale_2024} and time series prediction \citep{fujii_harnessing_2017,mujal_time_2023,steinegger_predicting_2025}.
We demonstrate that using an optimal measurement operator enables systematic improvements of QRC performance in these tasks.
Besides considering QELMs, which employ a memoryless quantum feature map and are thus close to untrained conventional quantum ML models \footnote{By ``conventional'', we mean quantum ML models that have trainable internal layers instead of only a final trainable classical regression layer.}, we show that our method can also be applied to recurrent QRCs with internal memory. 
The latter systems are typically used for time series processing \citep{fujii_harnessing_2017,fujii_quantum_2020}
and we show that the kernel framework underlying supervised quantum models naturally extends to the recurrent setting by constructing an exact Hilbert--Schmidt kernel on history space.

\section{Theory}
\label{sec_theory}

\begin{table}[tbp]
    \centering
    \begin{tabular}{l l}
        \hline
        Symbol & Meaning \\
        \hline
        $\Hcal$ & Hilbert space \\
        $\Herm(\Hcal)$ & Space of Hermitian operators on $\Hcal$ \\
        $N$ & Total number of qubits \\
        $N_I$, $N_A$ & Number of input/ancilla qubits \\
        $D$ & Hilbert space dimension ($D=2^N$) \\
        $d$ & Input data dimension ($\xv \in \reals^d$)$^{\star}$ \\
        $d'$ & Time series dimension ($\zv_t \in \reals^{d'}$)$^{\star}$ \\
        $P$ & Number of training samples \\
        $L$ & QRC memory length \\
        $T$ & Time series length \\
        $\numM$ & Number of measurement operators \\
        \hline
        $\xv$ & Input vector $^\star$ \\
        $\zv_t$ & Time series value at time $t$ $^\star$ \\
        $\yv$ & Target vector \\
        $\rho(\xv)$ & Density matrix of the quantum state (feature map) \\
        $\rho(x_t)$ & D.m.\ after evolving over the input history [\cref{eq_eff_seq_input}]\\
        $\eta_I$, $\eta_M$ & D.m.\ of the input and memory states \\
        $f(\xv)$ & Classical readout function \\
        $M_k$ & Measurement operators \\
        $w_k$ & Readout weights \\
        $K(\xv, \xv')$ & Quantum kernel \\
        $\alpha^*$ & Optimal kernel coefficients \\
        $M^*$ & Optimal measurement operator \\
        $P_k$ & $N$-qubit Pauli operators \\
        $\sigma^{j}$ & Pauli matrices ($j\in\{0,x,y,z\}$) \\
        \hline
    \end{tabular}
    \caption{Summary of notation used in this work. $^\star$For QELMs, $\xv=\zv_t$ and $d=d'$, while for stateful QRCs, the identification in \cref{eq_eff_seq_input} is necessary.}
    \label{tab:notation}
\end{table}

\subsection{Model}
\label{sec_gen_model}

We consider an $N$-qubit model, with dimension $D=2^N$ of its Hilbert space $\Hcal$, described by the feature map
\beq \xv\mapsto\rho(\xv),
\label{eq_qrc_feature_map}\eeq
which encodes input data $\xv\in\reals^d$ into a quantum state represented by a density matrix $\rho(\xv)$.
The density matrix is an element of the ($D^2=4^N$)-dimensional space $\Herm(\Hcal)$ of Hermitian operators acting on $\Hcal$ and can include the action of various quantum channels (such as unitary time evolution or noise) on some initial state $\rho_0$ (see \cref{sec_qrc_model} for specific examples).
The classical output of a QRC is given by 
\beq f(\xv) = \sum_{k=1}^\numM w_k \tr[M_k \rho(\xv)] ,
\label{eq_QELM}\eeq 
where $\{M_k\}$ are a set of chosen measurement operators (total number $\numM$) and $w_k$ are trainable weights.
We discuss the optimization problem in detail in the next section.

Importantly, for the present purpose, the feature map \eqref{eq_qrc_feature_map} is assumed to be \emph{stateless} in the sense that the state $\rho$ is always fully determined by the input vector $\xv$. 
Such a model is naturally realized by a \emph{quantum extreme learning machine} (QELM) \citep{fujii_quantum_2020,mujal_opportunities_2021,sakurai_quantum_2022}.
This seems at first to preclude QRCs with internal \emph{memory}, as commonly used for time series processing \citep{fujii_harnessing_2017,fujii_quantum_2020,mujal_opportunities_2021}, because their current state $\rho(t)$ would not only depend on the present input, but also on the inputs preceding it. To address this issue, let $\{\zv_0, \zv_1, \ldots, \zv_T\}$ be a time series ($T+1$ samples in $\reals^{d'}$) and assume that the QRC has a memory of length $L$. This approximation is typically warranted due to the echo state property of a QRC. Then, an effective input $x_t$ at time $t\geq L$ can be defined as 
\beq x_t := (\zv_{t-L}, \ldots, \zv_t) .
\label{eq_eff_seq_input}\eeq
This quantity can be understood as an ordered set of $L+1$ samples $\zv_{t-L}, \ldots, \zv_t$ such that processing these samples through the QRC is sufficient to obtain the unique state $\rho(t)$ associated with $x_t$.
Accordingly, the QRC output is fully determined by any of the $P=T-L$ unordered effective inputs $\{x_t\}_{t=L}^{T-1}$ and can thus be considered stateless in this setting \footnote{In practice, an initial washout/equilibration time for the QRC must be considered, which, however, does not affect the present theoretical framework.}. 
Prescriptions like \eqref{eq_eff_seq_input} to define truncated input histories are also used in kernelization of classical reservoir computing models \citep{gonon_reservoir_2022,grigoryeva_infinitedimensional_2025}.

\subsection{Measurement operator optimization}
\label{sec_msmtopt}

\begin{table}[t]
\centering
\small
\renewcommand{\arraystretch}{1.25}
\begin{adjustwidth}{-2.0cm}{-1.5cm}
\begin{tabular}{p{0.18\linewidth} p{0.27\linewidth}| p{0.27\linewidth}| p{0.27\linewidth}}
\hline
& \textbf{(A) Fixed measurements (constrained primal)}
& \textbf{(B) Free observable (unconstrained primal)}
& \textbf{(C) Kernel form (dual of B)} \\
\hline\hline

Given data
& \multicolumn{3}{p{0.81\linewidth}}{$\{(\mathbf{x}_i,y_i)\}_{i=1}^P$, reservoir states $\rho_i:=\rho(\mathbf{x}_i)\in\Herm(\Hcal)$} \\
\hline

Trainable object
& $\mathbf{w}\in\mathbb{R}^\numM$ 
& $M\in\Herm(\Hcal)$
& $\boldsymbol{\alpha}\in\mathbb{R}^P$ \\
\hline

Allowed observables
& $M=\sum_{k=1}^\numM w_k M_k$\newline $\in S:=\mathrm{span}\{M_k\}$
& any $M\in\Herm(\Hcal)$
& $M=\sum_{i=1}^P \alpha_i \rho_i$ (representer form) \\
\hline

Prediction
& $f(\mathbf{x})=\sum_{k=1}^\numM w_k \tr (M_k\rho(\mathbf{x}))$
& $f(\mathbf{x})=\tr (M\rho(\mathbf{x}))$
& $f(\mathbf{x})=\sum_{i=1}^P \alpha_i K(\mathbf{x}_i,\mathbf{x})$ \\
\hline

Training objective
& $\displaystyle \min_{\mathbf{w}} \onehalf \sum_{i=1}^P \Bigl(y_i-\sum_{k=1}^\numM w_k \tr (M_k\rho_i)\Bigr)^2 + \frac{\lambda}{2} \|\mathbf{w}\|_2^2$
& $ \min_{M} \onehalf \sum_{i=1}^P \bigl(y_i-\tr (M\rho_i)\bigr)^2 + \frac{\lambda}{2}\, \tr (M^2)$
& $\displaystyle \min_{\boldsymbol{\alpha}} \onehalf \|\mathbf{y}-\mathbf{K}\boldsymbol{\alpha}\|_2^2 + \frac{\lambda}{2}\, \boldsymbol{\alpha}^\top \mathbf{K}\boldsymbol{\alpha}$ \\
\hline

Matrices 
& $\Phi_{ik}:=\tr (M_k\rho_i)$
& $K_{ij}:=\tr (\rho_i\rho_j)$
& $K_{ij}:=\tr (\rho_i\rho_j)$ \\
\hline

Solution
& $\mathbf{w}^\star=(\Phi^\top\Phi+\lambda\mathbb{I})^{-1}\Phi^\top \mathbf{y}$
& $M^\star=\sum_{i=1}^P \alpha_i^\star \rho_i$
& $\boldsymbol{\alpha}^\star=(\mathbf{K}+\lambda\mathbb{I})^{-1}\mathbf{y}$  \\
\hline

Corresponding kernel
& $K_{\mathrm{meas}}(\mathbf{x},\mathbf{x}')=\boldsymbol{\varphi}(\mathbf{x})^\top\boldsymbol{\varphi}(\mathbf{x}')$,
$\varphi_k(\mathbf{x})=\tr (M_k\rho(\mathbf{x}))$
& $K(\mathbf{x},\mathbf{x}')=\tr (\rho(\mathbf{x})\rho(\mathbf{x}'))$
& $K(\mathbf{x},\mathbf{x}')=\tr (\rho(\mathbf{x})\rho(\mathbf{x}'))$ \\
\hline

Equivalence statement
& not equivalent to (B)/(C) unless $S$ is complete on the data manifold (no information loss)
& exactly equivalent to (C) (same objective, different parametrization)
& exactly equivalent to (B) (same objective, different parametrization) \\
\hline

Meaning of ``optimal''
& minimizer over $M\in S$ (best within chosen measurement family)
& minimizer over all $M\in\Herm(\Hcal)$ (best linear readout)
& minimizer in RKHS induced by $K$ (same solution as B) \\

\hline
\end{tabular}
\end{adjustwidth}
\caption{Three training formulations for a QRC. The training objective and solution are specialized here to the case of least-squares loss and assuming an orthonormal operator basis. For other kinds of losses, usually no analytic expressions for the weights $\wv^*$ exist. If the QRC carries an internal memory state, the above formulations remain intact but the data $\xv_i \hat\equiv x_t$ represents an input history [see \cref{eq_eff_seq_input}] and $\rho(\xv) \hat\equiv \rho(x_t)$ is the state resulting by sequentially processing that history (see \cref{sec_qrc_dynamics,app_qrc_stateful}).}
\label{tab:qrc_opt_forms}
\end{table}

When training a RC-type model, there are in principle three formalisms available: (A) constrained primal  optimization, (B) unconstrained primal optimization, and (C) dual space optimization \citep{bishop_pattern_2006}. Method (C) seems not to have been discussed previously in the context of QRCs.
\paragraph*{Constrained primal optimization.} 
The conventional way of training a QELM like in \cref{eq_QELM} consists in selecting a set of measurement operators $\{M_k\}$ and optimizing the weights $w_k$ via regression on a training data set $\{(\xv_i,y_i)\}_{i=1}^P$ of $P$ samples:
\beq \min_{\mathbf{w}\in\reals^\numM} \sum_{i=1}^P \Lcal \left( \sum_{k=1}^\numM w_k \tr[M_k \rho(\xv_i)],\, y_i \right) + \frac{\lambda}{2} \|\mathbf{w}\|^2_2 ,
\label{eq_primal_opt}\eeq
where $\Lcal$ is a loss function and $\lambda$ is a regularization parameter.
For regression and time-series prediction tasks, one typically uses the least-squares loss $\Lcal(\hat y,y)=\frac{1}{2}(\hat y-y)^2$, in which case the optimal weights are given by
\beq \mathbf{w}^* = (\boldsymbol{\Phi}^\top \boldsymbol{\Phi} + \lambda \Imat)^{-1} \boldsymbol{\Phi}^\top \mathbf{y}, \qquad \Phi_{ik} = \tr(M_k \rho(\xv_i)),
\label{eq_primal_weights}\eeq
where $\mathbf{y}\in\reals^P$ is the sample vector of target values and $\varphi_k(\xv)=\tr(M_k \rho(\xv))$ can be considered as the feature map.
For classification tasks, one often uses the hinge loss $\Lcal\st{hinge}(\hat y,y)=\max(0,1-y\hat y)$ (with $y\in\{-1,+1\}$), in which case the resulting optimal weights then generally do not have a closed-form expression. For multi-class classification with $C$ classes, a common strategy is to train $C$ separate readouts (one-vs-rest), i.e., solve \cref{eq_primal_opt} for each class $c=1,\ldots,C$ with targets $y_i^{(c)}\in\{-1,+1\}$ and separate weights $\mathbf{w}^{(c)}$.

\paragraph*{Unconstrained primal optimization.}
Crucially, unless the $\{M_k\}$ span the full operator space $\Herm(\Hcal)$, the optimization in \cref{eq_primal_opt} does not guarantee that all relevant information encoded in the quantum state $\rho$ can actually be accessed. The resulting trained QRC may thus perform poorly on the given data set.
In variational QML, this issue is addressed by training parametrized unitaries acting on $\rho$ before performing the measurement.
Alternatively, and specifically for QRCs, we can exploit the freedom in the choice of measurement operators and hence optimize over the \emph{full} Hermitian operator space $\Herm(\Hcal)$: 
\beq \min_{M\in\Herm(\Hcal)} \sum_{i=1}^P \Lcal\bigl( \tr[M \rho(\xv_i)],\, y_i \bigr) + \frac{\lambda}{2} \tr(M^2)
\label{eq_primal_opt_unc}\eeq
with the same loss function $\Lcal$ as in \cref{eq_primal_opt}.
In practice, this optimization over $\Herm(\Hcal)$ can be achieved by using in \eqref{eq_primal_opt} as observables $\{M_k\}$ a complete orthonormal $\numM=D^2$-dimensional basis of $\Herm(\Hcal)$ (e.g., the normalized Pauli basis $\{2^{-N/2} P_k\}$). The associated optimal weights $w_k^*$ are given formally by the same expression as in \cref{eq_primal_weights}, and yield the optimal observable 
\beq M^* = \sum_{k=1}^{D^2} w_k^* M_k.
\label{eq_Mopt_primal}\eeq
For the least-squares loss, the solution weights $w_k^*$ are given by \cref{eq_primal_weights}. 

\paragraph*{Dual optimization.}
The above unconstrained optimization procedure becomes especially natural in its dual formulation, resting on the connection between supervised quantum ML models and quantum kernels \citep{schuld_supervised_2021}.
Identifying the unconstrained feature map as $\phi(\xv)=\rho(\xv)$ allows one to define the kernel $K(\xv,\xv')=\tr(\rho(\xv)\rho(\xv'))$, which induces a reproducing kernel Hilbert space (RKHS) \citep{scholkopf_generalized_2001}.
The representer theorem implies that the optimizer $M^*$ (\emph{optimal measurement operator}) of the problem \eqref{eq_primal_opt_unc} is of the form (see also \cref{app_qelm_kernel})
\beq M^* = \sum_{j=1}^P \alpha_j^* \rho(\xv_j),
\label{eq_Mopt}\eeq  
i.e., it necessarily sits in the span of the training features \citep{scholkopf_generalized_2001,schuld_supervised_2021}.
In terms of the representer coefficients $\boldsymbol{\alpha}\in \reals^P$, the corresponding dual problem reads
\beq \min_{\boldsymbol{\alpha}\in\reals^P} \sum_{i=1}^P \Lcal\bigl( (\mathbf{K}\boldsymbol{\alpha})_i,\, y_i \bigr) + \frac{\lambda}{2}\, \boldsymbol{\alpha}^\top \mathbf{K}\boldsymbol{\alpha}.
\label{eq_dual_opt}\eeq
For the least-squares loss, the canonical solution coefficients are available in closed form as
\beq \boldsymbol{\alpha}^* = (\mathbf{K} + \lambda \Imat)^{-1}\mathbf{y}, \qquad K_{ij} = \tr(\rho(\xv_i)\rho(\xv_j)).
\label{eq_opt_kernel_params}\eeq 
For other loss functions, the solution coefficients must typically be computed numerically.
Independently of the loss function $\Lcal$, an optimally trained QELM [\cref{eq_QELM}] admits the kernel form 
\beq f(\xv)=\sum_{j=1}^P \alpha_j^* \tr(\rho(\xv_j)\rho(\xv)) = \tr(M^* \rho(\xv))
\label{eq_QELM_kernel}\eeq
The generalization to multi-dimensional target values $\yv_i\in \reals^C$ ($i=1,\ldots,P$) is straightforward and requires a separate kernel construction like in \cref{eq_QELM_kernel} for each class (this includes multi-class classification in a one-vs-rest scheme).

When the measurement operators $\{M_k\}$ form a complete orthonormal basis (or at least span the relevant subspace containing $M^*$ \citep{innocenti_potential_2023}), primal and dual formulations are \emph{equivalent} and \cref{eq_Mopt_primal,eq_Mopt} describe the same object. Expanding the optimal observable $M^*$ in a complete orthogonal basis leads to a relation between the coefficients $\wv^*$ and $\boldsymbol{\alpha}^*$ via [cf.\ \cref{eq_primal_dual_weights,eq_elm_wts_sol_dual}]
\beq w_k^* = \frac{1}{g_k} \sum_{j=1}^P \alpha_j^* \tr(M_k \rho(\xv_j)). 
\label{eq_wk_dual} \eeq
Here, $\tr(M_k M_l) = g_k \delta_{kl}$ with $g_k = 2^N$ for the orthogonal Pauli basis.
The optimization formulations are summarized in \cref{tab:qrc_opt_forms}.

\paragraph*{Remarks.}
(i) Neither the primal nor the dual optimization methods assume uncorrelated samples. Both are therefore compatible with the sliding window construction \eqref{eq_eff_seq_input}. (ii) Global conjugation $\rho(\xv)\mapsto U\,\rho(\xv)\,U^\dag$ leaves the expression \eqref{eq_QELM_kernel} unchanged and only transforms the optimal observable as $M^*\mapsto U M^* U ^\dag$. (iii) The constrained primal formulation can be ``dualized'', too, by using the kernel induced by the constrained feature map (see \cref{tab:qrc_opt_forms}). While this does not render a lower training error, it may offer computational benefits.

\subsection{Enhancements and practical considerations}
\label{sec_enhancements}

Using the Hermitian $2^N\times 2^N$ matrix $M^*$ directly as a measurement operator can be  infeasible in a practical setting with large qubit numbers $N$. 
After addressing below the computational efficiencies of the primal vs.\ dual method, we consider possible approaches to reducing the implementation complexity.

\subsubsection{Efficiency of primal vs.\ dual optimization}
\label{sec_prim_dual}

When taking the complete $m=D^2=4^N$-dimensional basis of measurement operators $\{M_k\}$ in \cref{eq_QELM}, the primal method is equivalent to the dual one, and the preference for either method becomes a matter of computational efficiency:

\begin{itemize}
    \item The \emph{primal} formulation requires solving for $\numM$ weights, involving the inversion of a $\numM \times \numM$ matrix (covariance matrix) with time complexity $\Ocal(\numM^3)$ and memory complexity $\Ocal(\numM^2)$. In the case of a complete operator basis, $\numM=D^2=4^N$, leading to a scaling of $\Ocal(64^N)$ in time and $\Ocal(16^N)$ in memory.
    \item In contrast, the \emph{dual} formulation solves for $P$ coefficients by inverting the $P \times P$ kernel matrix, which scales as $\Ocal(P^3)$ in time and $\Ocal(P^2)$ in memory.
\end{itemize}
Thus, the dual method is preferable when the number of samples is small compared to the feature space dimension ($P \ll \numM$), which is typical for large qubit numbers $N$ due to the exponential scaling of $D$. If one chooses to truncate the operator basis to $m<D^2$, the primal method can become more efficient, however at the expense of losing the optimality guarantee.

Note that the resulting optimal measurement operator $M^*$ [\cref{eq_Mopt}] is still a $D \times D$ matrix. Explicitly constructing and storing $M^*$ requires $\Ocal(D^2)=\Ocal(4^N)$ memory, which eventually limits the feasibility of the full optimization for large $N$.
Conversely, if the number of measurement operators is small (e.g., due to truncation) or the dataset is very large such that $P \gg \numM$, the primal formulation becomes more efficient.
Specialized classical and quantum algorithms for the inversion of large matrices exist that can mitigate the exponential scaling (see, e.g., \cite{bukru_hybrid_2025}).

We finally remark that, in practice, one does not have to explicitly construct the samples as in \cref{eq_eff_seq_input} for the optimization in the stateful case.
Instead, one can simply run the QRC over the time series $\zv_t$ and (after some equilibration period) collect the output states $\{\rho_t\}_{t=L}^{T-1} \hat= \{\rho(\xv_j)\}_{j=1}^P$ or a subset of them. The sliding window construction is implicit in this procedure.

\subsubsection{Pauli basis projection}
\label{sec_pauli_proj}

In this approach, the Hermitian operator $M^*$ is expanded in the Pauli (or any other operator) basis,
\beq M^* = \sum_{k=0}^{4^N-1} c_k P_k,\qquad c_k = \frac{1}{2^N} \tr(M^* P_k),
\label{eq_pauli_exp}\eeq
where the $P_k$ are all possible tensor products of $N$ Pauli operators (including the identity operator) and the coefficients $c_k$ follow from orthogonality of the Pauli basis. 
The absolute values $|c_k|$ indicate the importance of the observable $P_k$ for the given prediction task.
Selecting only the most relevant observables or restricting to a subset of a given maximum Pauli weight renders a set of implementable observables $\Ocal\st{Pauli}^*=\{P_k\}_{k=1}^{K}$.
The resulting prediction error is discussed in \cref{app_truncation_error}.

A problem with this approach is that the elements of the set $\Ocal\st{Pauli}^*$ do not necessarily mutually commute, requiring in principle multiple executions of the quantum circuit to obtain the full prediction. 
Moreover, the Pauli expansion is highly redundant, as we will see in the following.

\subsubsection{Diagonalization}
\label{sec_diag}
Since $M^*$ is Hermitian, it can be diagonalized via a unitary $V$ as
\beq
M^* = V \Lambda V^\dag,\qquad \Lambda = \sum_{\bv{b}\in\{0,1\}^N} \lambda_\bv{b}\, |\bv{b}\ket\bra \bv{b}|,
\label{eq_diag_spec}
\eeq
where $\{|\bv{b}\ket\}_{\bv{b}\in\{0,1\}^N}$ is the computational basis and $\lambda_\bv{b}\in\reals$ are eigenvalues. For any state $\rho(\xv)$,
\beq
\tr(M^*\rho(\xv)) = \tr\!\big(\Lambda\, V^\dag \rho(\xv) V\big) = \sum_{\bv{b}\in\{0,1\}^N} \lambda_\bv{b}\, \bra \bv{b}| V^\dag \rho(\xv) V |\bv{b}\ket.
\label{eq_diag_expect}
\eeq
The prediction is the weighted average of the observed bitstrings using the eigenvalues $\{\lambda_\bv{b}\}$.
Thus, the optimal observable can be implemented by a single basis rotation $V^\dag$ followed by a standard computational-basis measurement. 
This is equivalent to measuring all $2^N$ distinct Pauli-$Z$ strings $\{Z_\sv\}_{\sv\in\{0,1\}^N}$, since (see \cref{sec_diag_remarks}) 
\beq \Lambda = \sum_{\sv\in\{0,1\}^N} \beta_\sv\, Z_\sv
\eeq 
with $\beta_\sv$ given by \cref{eq_walsh_hadamard}. 

The unitary $V$ may be treated as part of the readout (applied once just before measurement) or absorbed into the reservoir by redefining the last layer as $U_R \mapsto V^\dag U_R$. This global conjugation leaves the kernel $K(\xv,\xv')$ and the trained predictor unchanged (see \cref{sec_msmtopt}) and only affects the hardware realization of the optimal measurement.
While, strictly, a pre-rotation followed by computational-basis measurement is sufficient and equivalent to measuring the exact $M^*$, we will include in our numerical experiments [\cref{sec_numexp}] also the Pauli observable decomposition [\cref{eq_pauli_exp}], as both approaches can lead to slight differences in numerical stability.


\section{QRC models}
\label{sec_qrc_model}

The optimization methods described above are general and independent of the encoding scheme or other quantum channels acting on the readout state $\rho$.
In order to test our methods in practice, we consider some specific QRC models.

\subsection{QRCs and QELMs}
\label{sec_qrc_dynamics}

The standard (stateful) QRC model introduced in \cite{fujii_harnessing_2017,fujii_quantum_2020} consists of a set of coherently evolving (memory/ancilla) qubits ($N_A$) periodically coupled to a set of input qubits ($N_I$). 
At time $t$, a classical input $\zv_t \in \mathbb{R}^{d'}$ is encoded into an input state with density matrix $\eta_I(\zv_t)$. The joint state with the coherent ancilla state $\eta_M(t)$ is formed and time-evolved by a unitary $U$:
\beq \eta(t) = U \left(\eta_I(\zv_t) \otimes \eta_M(t)\right) U^\dagger.
\label{eq_qrc_rho}\eeq 
This $\eta(t)$ represents the feature map $\rho(x_t)$ on the input sequence defined by \cref{eq_eff_seq_input} [corresponding to the one in \eqref{eq_qrc_feature_map}].
Subsequently, the input is reset and discarded, and the memory alone is carried to the next step:
\beq 
\eta_M(t+1)  = \tr_I\!\left[\eta(t)\right],
\label{eq_qrc_update}\eeq
where $\tr_I$ denotes partial trace over the input register $I$ (see \cref{app_qrc_stateful} for further details).

We assume the encoding state to be pure, $\eta_I(\zv)=|\psi(\zv)\ket\bra \psi(\zv)|$, with the state vector $|\psi(\zv)\ket \in \Hcal_I$ (living in the $N_I$-qubit input Hilbert space) given by 
\beq |\psi(\zv)\ket = S(\zv) |\Phi_0\ket
\label{eq_q_feature_map}\eeq 
in terms of the encoding unitary $S(\zv)$ and the initial state $|\Phi_0\ket=|0\ket^{\otimes N_I}$. 
The \emph{reservoir} unitary $U$ is defined by gate operations or Hamiltonian time evolution, e.g., $U(t)=\exp(-\im t H_R)$ with a reservoir Hamiltonian $H_R$. 
Specifically, we consider a transverse field Ising Hamiltonian with coupling parameters $h$ and $J$:
\beq H_R = -h \sum_{j=1}^N \sigma_j^x - J \sum_{j=1}^{N-1} \sigma_j^z \sigma_{j+1}^z,
\label{eq_TFIM_H}\eeq
as well as the variant with $\sigma^x$ and $\sigma^z$ exchanged.
Here, $\sigma^p_j$, $p\in \{0,x,y,z\}$ ($\sigma^0_j\equiv \Imat$) denote single-qubit Pauli matrices acting on the $j$-th qubit.

For a \emph{QELM}, the coherent qubits are absent ($N_A=0$) and \cref{eq_qrc_rho} reduces to a stateless feature map
\beq \rho(\xv)=\eta(\xv)=U\eta_I(\xv) U^\dag
\label{eq_qelm_rho}\eeq
without time evolution.
The sliding window prescription \eqref{eq_eff_seq_input} is not relevant in this case and one can take $\xv=\zv_t$ to process time series.
Moreover, in this setting, any data-independent unitary drops out from the trained predictor in \cref{eq_QELM_kernel} as 
\beq
K(\xv,\xv')=\tr\big(\rho(\xv)\rho(\xv')\big)= \tr(\eta_I(\xv)\eta_I(\xv')) =\big|\bra\Phi_0| S^\dagger(\xv) S(\xv') |\Phi_0\ket\big|^2.
\label{eq_kernel_invar}\eeq
This does not hold for the QRC dynamics \eqref{eq_qrc_update} with memory.

\subsection{Encoding schemes}
\label{sec_encodings}

The encoding $S(\zv)$ is a crucial component of any quantum ML model.
In our numerical experiments, we use parallel product state encodings of the form \citep{schuld_effect_2021}
\beq
S(\zv)=\bigotimes_{j=1}^{N_I} S_j(z_j),
\label{eq_encoding_S}\eeq
with the unitary operator $S_j$ encoding a single scalar $z_j$ (thus $N_I=d$) and acting on the $j$-th qubit.
We consider the following specific encoding schemes.

(1) \emph{Rotational} encoding:
\beq
S_j(z_j) = R_{p_j}(2\pi z_j),
\label{eq_encop_rot}\eeq
where
\beq R_p(\theta)\equiv \exp\left(-i \frac{\theta}{2} \sigma^p\right) =\cos\left(\frac{\theta}{2}\right)\Imat - i\sin\left(\frac{\theta}{2}\right)\sigma^p
 \label{eq_rot_gates}\eeq 
are standard rotation gates. 
We will mostly use $R_Y$ encoding ($p=y$) for all qubits, which results in the state
\beq
 |\psi_{\text{rot}}(z)\ket = R_Y(2\pi z)|0\ket = \cos(\pi z)|0\ket+\sin(\pi z)|1\ket.
\label{eq_enc_rot_Y}\eeq
This implies a useful data scale of $z \in [0, 1]$, since $z \in [1, 2]$ produces only an irrelevant global phase relative to $z \in [0, 1]$.
An $R_X$ gate acts similarly, but introduces a complex phase, which is undesired and not needed for classical data processing.
$R_Z$ gates alone are not suitable for encoding as they just shift the phase of the $|0\ket$ state. 
They can, however, be meaningfully combined with $R_Y$ gates to encode two scalars into the same qubit (dense encoding): 
\beq \widehat S_j(z_j,z_{j+1}) = R_Z(2\pi z_{j+1}) R_Y(2\pi z_j),
\label{eq_enc_rot_dense}\eeq
where now $N_I=d/2$ \citep{larose_robust_2020,wang_development_2022}.

(2) \emph{Amplitude} encoding: inputs are scaled to $z\in[0,1]$ and prepared as single-qubit states that traverse half a great circle on the Bloch sphere:
\beq
|\psi\st{amp}(z)\ket = \sqrt{z}|0\ket + \sqrt{1-z}|1\ket .
\label{eq_enc_amp_sqrt}\eeq
A symmetric variant maps $z\in[-1,1]$ to the full great circle on the Bloch sphere:
\beq
|\psi\st{amp'}(z)\ket = z|0\ket+\sqrt{1-z^{2}}|1\ket .
\label{eq_enc_amp_sq}\eeq
We remark that these encodings can be expressed in terms of single $R_Y$ rotations, e.g., $|\psi\st{amp'}(z)\ket = z|0\ket + \sqrt{1-z^{2}}|1\ket = R_Y(2\arccos z)|0\ket$.

\subsection{Measurement readout}
\label{sec_readout}
In practice, the measured expectation values in \cref{eq_QELM} can be augmented by nonlinear features in order to enhance the expressivity of the model \citep{yang_design_2018,pauwels_photonic_2021,zhu_minimalistic_2024,steinegger_predicting_2025}. 
Defining the vector of measurement outcomes
\beq
\vv(\xv) := (v_1(\xv),\ldots,v_\numM(\xv))^\top,\qquad v_k(\xv)=\tr[M_k \rho(\xv)],
\eeq
we construct a readout vector $\Rv(\xv)$ consisting of all distinct monomials in the components of $\vv(\xv)$ of total degree up to $p\st{max}$. 
Introducing the  multi-index $\boldsymbol{\alpha}\in\mathbb{N}_0^\numM$ with $|\boldsymbol{\alpha}| := \sum_{k=1}^\numM \alpha_k$,
we can write 
\beq
\Rv(\xv) := \left(\prod_{k=1}^\numM v_k(\xv)^{\alpha_k}\right)_{1\leq |\boldsymbol{\alpha}|\leq p\st{max}} \in \reals^{m\st{ro}},
\label{eq_QELM_readout}\eeq
where any fixed ordering of the multi-indices $\boldsymbol{\alpha}$ specifies the components $R_j$ of $\Rv$. The number of distinct monomials of total degree $p$ equals $\binom{\numM+p-1}{p}$, so that the total readout dimension is
\beq
m\st{ro} = \sum_{p=1}^{p\st{max}} \binom{\numM+p-1}{p} = \binom{\numM+p\st{max}}{p\st{max}}-1.
\eeq
Taking into account $C$ possible output channels/classes with targets $\yv_i=(y_i^{(1)},\ldots,y_i^{(C)})^\top\in\reals^C$, the complete QRC output function becomes
\beq
\fv(\xv) = \Wv^\top \Rv(\xv),\qquad \Wv=[\wv^{(1)},\ldots,\wv^{(C)}]\in\reals^{m\st{ro}\times C}.
\label{eq_QELM_output}\eeq
Note that the measurement-operator optimization theory developed here applies only to the linear readout case.

\section{Numerical experiments}
\label{sec_numexp}

In the following, we present numerical experiments on QRCs and QELMs trained using the kernel-based measurement operator optimization.
We will typically show test accuracy as a function of the number of measurement operators kept after thresholding.
The maximum number of operators resulting from the diagonalization and Pauli projection methods [\cref{sec_pauli_proj,sec_diag}] is $\numM\st{diag}=2^N\times D\st{out}$ and $\numM\st{pauli}=4^N\times D\st{out}$, respectively, where $D\st{out}$ is the output dimension. 
In the case of setups without reservoir unitaries, the effective maximum number can be lower, since expectation values involving Pauli-$Y$'s vanish identically on states encoded via \eqref{eq_enc_rot_Y}, \eqref{eq_enc_amp_sqrt}, or \eqref{eq_enc_amp_sq} (see \cref{app_prodstates}).
After having determined the optimal operator set, we generally re-fit the QRC using this set [alternatively, one may use \cref{eq_wk_dual}].

\subsection{Image classification task} 

We first consider the standard QELM task of classifying handwritten digits from the MNIST dataset \citep{lecun_handwritten_1989}, which contains about $P\simeq 70\:000$ images of size 28$\times$28 pixels \citep{sakurai_quantum_2022,kharsa_advances_2023,lorenzis_harnessing_2025}.
We use a QELM with $N=5$ qubits and dimensionally reduce the images to their $2N$ dominant principal components, which we encode via $R_Y$-$R_Z$ rotation operators (allowing for two features per qubit). 
For the kernel-based optimization, we use a random subset of approximately 10\:000 samples.
For comparison to a classical model, we consider an SVM with polynomial features (up to degree 4) constructed from the principal components.

\begin{figure}[tb!]
    \centering
    \includegraphics[width=0.48\textwidth]{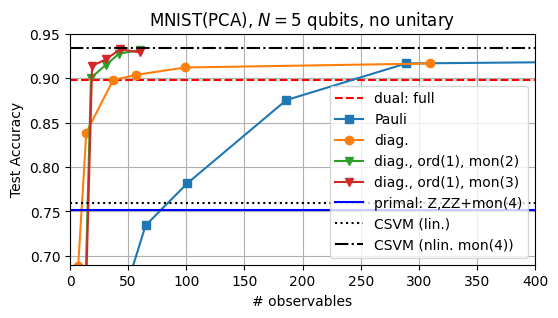}\quad
    \includegraphics[width=0.48\textwidth]{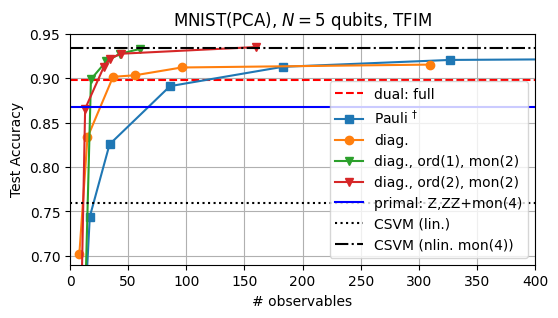}
    \caption{Optimization of the measurement operator for a QELM trained on the MNIST (first $2N$ principal components) classification task (left) without and (right) with a reservoir unitary (transverse field Ising model). The dashed line represents the test accuracy obtained using the full operator $M^*$ (i.e., one per digit class), while the connected symbols correspond to the Pauli decomposition and diagonalization approaches [\cref{eq_pauli_exp,eq_diag_spec}]. The notation ord($k$) and mon($p$) refers to the observable order $k$ and monomial order $p$ in \cref{eq_QELM_readout}. The curve labeled `primal: Z,ZZ+mon(4)' represents primal optimization using only the Pauli-$Z$ and -ZZ observables enhanced by readout monomials up to 4th order, while CSVM refers to a classical SVM trained on (non-)linear features constructed from the principal components. Since there is no intrinsic randomness in the model, error bars arise only from the selection of samples and are negligible here. $^\dag$Here, we optimized the kernel only over a single class (to limit combinatorial explosion in the number of readout terms), but used the resulting operator set to re-train the QELM for all classes. }
    \label{fig:qelm_opt_obs}
\end{figure}

The prediction accuracy on a held-out test data set ($P\st{test}=0.2 P$) is given by the ratio of correctly classified over total number of samples:
\beq
\Acal = \frac{1}{P\st{test}}\sum_{i=1}^{P\st{test}} \delta_{y_i, \hat y_i}, \quad \text{with } \hat y_i = \arg\max_c \hat f^{(c)}(\xv_i).
\label{eq_test_acc}
\eeq
\Cref{fig:qelm_opt_obs} shows the resulting test accuracies for the MNIST classification problem in the case without (left panel) and with (right panel) a reservoir unitary.

\paragraph*{QELM without a reservoir.} 
We find that the test accuracy of the classical SVM saturates at around $\Acal\st{CSVM}\approx 93.5\%$, which appears to be also an upper limit for the QELM. 
This limit is reached when the QELM readout is enhanced by monomials up to order $p\st{max}=2$ [see \cref{eq_QELM_output}], both for the Pauli-decomposition and diagonalization approaches [\cref{fig:qelm_opt_obs}(a)]. In this case, it suffices even to restrict the observables to weight-1 and -2 Pauli strings.
Notably, without additional monomial features, the QELM test accuracy saturates at $\Acal\st{QELM}\approx 92\%$, but requires a significantly larger number of observables.
This quantum-to-classical performance gap exists because powers $\langle \sigma_j \rangle^q$ of expectation values of observables acting on the \emph{same} qubit $j$ cannot be represented in the operator basis for the present encoding.

\paragraph*{QELM with a reservoir.}
\Cref{fig:qelm_opt_obs}(b) shows the test accuracies for a QELM in the same setup as above, but \emph{with} a reservoir unitary based on the TFIM \eqref{eq_TFIM_H} (with parameters $t h\approx t J \approx 10$).
When optimizing via the primal route using only $Z$- and $ZZ$-Pauli observables enhanced with monomials up to order $p\st{max}=4$, we obtain a higher test accuracy in the case with than without a reservoir unitary. We explain this by the information scrambling property of the unitary, which makes information that sits in the Pauli-$X$- and $Y$-sectors accessible in the $Z$ sector.
In general, the results obtained for the QELM optimized via the kernel method are similar both with and without a reservoir unitary, demonstrating the robustness of the approach.

We summarize this section with some common findings: (i) Unless one uses a complete operator basis or techniques like temporal multiplexing \citep{fujii_harnessing_2017}, the primal optimization is generally inferior to the kernel-based approach: in the latter, it suffices to restrict the readout to Pauli-$Z$ observables enhanced by 2nd-degree monomials to reach the classical SVM accuracy $\Acal\st{CSVM}$. The fundamental reason is of course that the kernel-based method implicitly renders an optimized pre-rotation before measurement.
(ii) Performing the kernel-based optimization only over a single class can significantly reduce the computational load without sacrificing accuracy, provided the QELM is retrained for all classes. (iii) Low order monomial in the readout (especially squares) are generally helpful to close the quantum-to-classical accuracy gap (depending on the specific encoding).

\subsection{Time series prediction}

We turn to autoregressive prediction tasks \citep{fujii_harnessing_2017,mujal_opportunities_2021,martinez-pena_dynamical_2021}, where, given a time series $\{\zv_t\}_{t=0}^T$, the QRC is trained to predict the next step $\hat \zv_{t+1} = f(x_t)$ (with the prescription \eqref{eq_eff_seq_input} where appropriate). 
A complete forecasted trajectory is generated by repeatedly applying the prediction function, i.e., $\hat \zv_{t+1} = f(\hat x_t)$.
As was the case for static data, enhancing the readout with monomials typically improves forecasting capability \citep{zhu_minimalistic_2024,steinegger_predicting_2025}. However, they obscure a clear assessment of the performance of the quantum parts of the model, and we therefore consider here only setups without hand-engineered features in the readout.
As a performance measure, we consider, beside the root-mean-square training error $E\st{train}=\left[\frac{1}{P}\sum_{t=0}^{P-1} \|\zv_{t+1} - f(\zv_t)\|^2\right]^{1/2}$, the forecast horizon 
\beq T\st{fch} = \inf \left\{ t : \| \zv(t) - \hat{\zv}(t) \| > \sigma \right\}
\label{eq_fc_hor}\eeq 
where $\sigma^2= \frac{1}{P}\sum_{i=0}^{P-1} \| \bar \zv - \zv(t_i)\|^2$ is the variance of the trajectory and $\bar \zv = \frac{1}{P}\sum_{i=0}^{P-1} \zv(t_i)$ is its mean. 
In the case of chaotic systems, the forecast horizon is expressed in units of the Lyapunov time $T_L=1/\lambda_1$ with largest Lyapunov exponent $\lambda_1$.

\begin{figure}[tb!]
    \centering
    \includegraphics[width=0.45\textwidth]{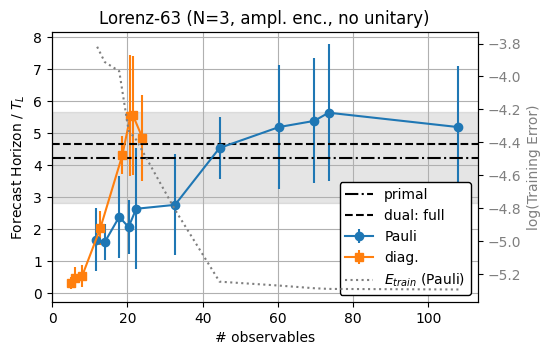}
    \includegraphics[width=0.45\textwidth]{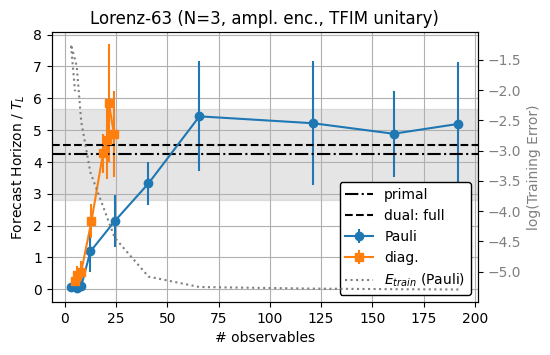}
    \caption{Optimization of the measurement layer for a memoryless QRC without ancilla qubits (QELM), for the cases \emph{without} and \emph{with} a (TFIM-based) reservoir unitary. The QRC is trained on a 3-dimensional chaotic time series [\cref{eq_Lorenz63}]. The error bars on the forecast horizon represent the standard deviation (not error of the mean) after averaging over different initial conditions. (Their magnitude is comparable to classical RC results for chaotic systems.) The dashed line represents the forecast horizon when using the kernel-based optimal measurement operator $M^*$ directly, while the connected points correspond to decompositions into subsets of observables [\cref{eq_pauli_exp,eq_diag_spec}]. The dash-dotted line gives the forecast horizon (with its standard deviation indicated by the gray area) for the primal optimization over the complete operator basis. $E\st{train}$ denotes the (RMS) training error. The horizontal axis represents the number of observables ranked by their relevance (left = least relevant; maximum number $= 64\times 3=192$ for right panel).}
    \label{fig:qrc_opt_free}
\end{figure}

\subsubsection{Memoryless QRCs (QELMs)}

Time series generated by Markovian dynamical systems can be learned using QELMs, i.e., memoryless QRCs (cf.\ \cite{ahmed_optimal_2024,ahmed_prediction_2024,mccaul_minimal_2025} for similar minimal QRC approaches and \cite{gross_flow_2026} for a discussion within a classical RC framework). 
We can thus train the QRC directly on the individual samples of the time series, $\xv_t=\zv_t$ and ignore the prescription \eqref{eq_eff_seq_input}.
As a standard benchmark for reservoir computing prediction tasks, we take the $(d=3)$-dimensional Lorenz-63 chaotic system, described by 
\beq 
\begin{aligned}
\dot z_1 &= \sigma (z_2 - z_1) \\
\dot z_2 &= z_1 (\rho - z_3) - z_2 \\
\dot z_3 &= z_1 z_2 - \beta z_3
\end{aligned}
\label{eq_Lorenz63}\eeq
with parameters $\sigma = 10$, $\rho = 28$, $\beta = \sfrac{8}{3}$ and largest Lyapunov exponent $\lambda_1\approx 0.89$ \citep{sprott_elegant_2010}.
The trajectory is numerically generated using a RK4 (Runge-Kutta) scheme with tolerances of $10^{-10}$.
We use the amplitude encoding \eqref{eq_enc_amp_sqrt} with $N_I=3$ qubits. We find that the other tested encoding schemes yield similar training accuracy.

The resulting forecast horizons are illustrated in \cref{fig:qrc_opt_free} for the case without (left) and with a reservoir unitary (right). 
Due to using an optimal measurement operator $M^*$, the presence of a reservoir unitary does not have a significant impact on the largest achievable forecast horizon. 
Interestingly, both for the Pauli-decomposition and the diagonalization approach, the largest forecast horizon is reached for a slightly reduced number of observables. We explain this by the numerical conditioning effects and redundancy in the operator basis for the present dataset and encoding scheme.

\subsubsection{QRCs with internal memory}

\begin{figure}[tb!]
    \centering
    \includegraphics[width=0.45\textwidth]{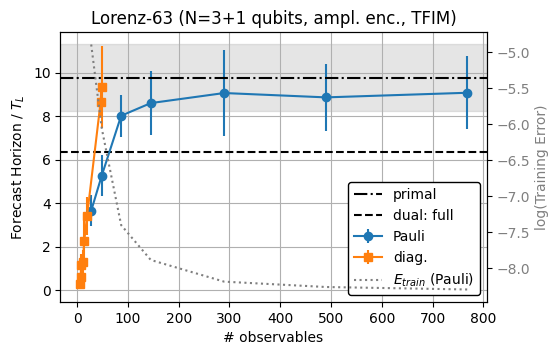}
    \caption{Optimization of the measurement operator for a QRC with internal memory ($N_A=1$ ancilla qubit coupled via a TFIM unitary to $N_I=3$ input qubits), trained to predict the Lorenz-63 time series [\cref{eq_Lorenz63}] encoded via amplitude encoding [\cref{eq_enc_amp_sqrt}]. The dashed line represents the forecast horizon using the optimal operator $M^*$ directly, while the connected points correspond to decompositions into subsets of observables [\cref{eq_pauli_exp,eq_diag_spec}]. The forecast horizon obtained from the primal optimization over the full operator basis is shown by the dashed-dotted line, with the gray area representing the standard deviation. }
    \label{fig:qrc_opt_anc_lor}
\end{figure}

We now turn to a \emph{stateful} QRC, obtained by extending the above QELM model by $N_A\geq 1$ ancilla qubits, coupled via a TFIM unitary. The measurement operator acts on all $N=N_I+N_A$ qubits and is optimized using the kernel-based approach.
We first revisit the forecasting problem for the Lorenz-63 model and then consider two strongly non-Markovian time series: a harmonic signal and the Mackey-Glass system.

\begin{figure}[tb!]
    \centering
    \includegraphics[width=0.46\textwidth]{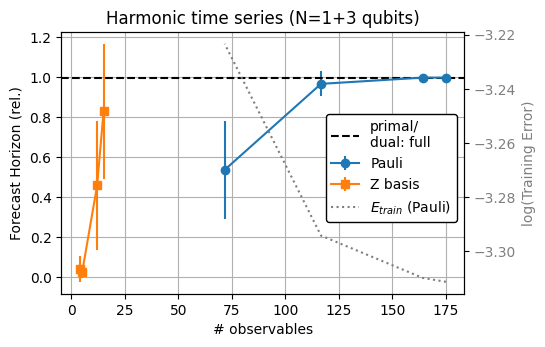}\qquad
    \includegraphics[width=0.48\textwidth]{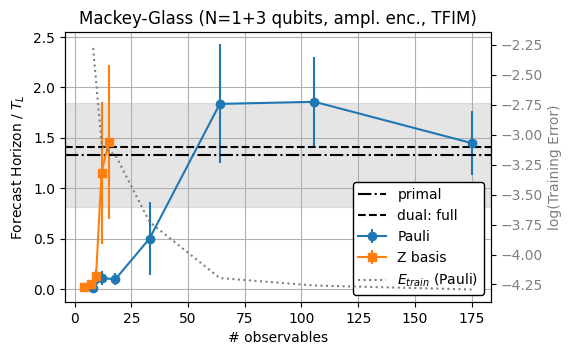}
    \caption[]{Optimization of the measurement layer for a QRC \emph{with} internal \emph{memory} ($N_A=3$ ancilla qubits, coupled to $N_I=1$ input qubits via a TFIM-based unitary). In (a), the QRC is trained on a random harmonic signal [\cref{eq_harm_timeser}] with $n=9$ frequencies, encoded via amplitude encoding \eqref{eq_enc_amp_sq}. In (b), the Mackey-Glass time series [\cref{eq_mcgl_DDE}] is encoded via \cref{eq_enc_amp_sqrt}. The dashed line represents the forecast horizon (relative to the maximum tested time in (a)) using the optimal measurement operator $M^*$ (as computed by the kernel-based optimization), while the connected points correspond to decompositions into subsets of observables (\cref{eq_pauli_exp,eq_diag_spec}, ranked by their relevance, along horizontal axis). The error bars on the forecast horizon represent the standard deviation after averaging over different initial conditions. The forecast horizon obtained from the primal optimization over the full operator basis is shown by the dashed-dotted line, with the gray area representing the standard deviation. $E\st{train}$ denotes the (RMS) training error.}
    \label{fig:qrc_opt_stateful}
\end{figure}

\paragraph*{Lorenz-63 time series.}
Using memory states significantly enhances prediction accuracy both for the primal and dual optimization method (\cref{fig:qrc_opt_anc_lor}), yielding forecast horizons of $T\st{fch}/T_L\approx 10$ (which is comparable to recent literature results \citep{steinegger_predicting_2025}).
Notably, using the optimal operator $M^*$ directly leads in this case to a significantly lower forecast horizon and higher training error ($E\st{train}\approx 2\times 10^{-6}$) compared to the Pauli decomposition or diagonalization methods ($E\st{train}\approx 4\times 10^{-8}$). 
This result appears to be specific to the current QRC setup or data, as we do not observe strong discrepancies in the other benchmarks (cf.\ \cref{fig:qrc_opt_free,fig:qrc_opt_stateful}). 

\paragraph*{Harmonic time series.}
A scalar harmonic time series consists of $n$ frequencies with random amplitudes $\{a_k, b_k\}$:
\begin{equation}
    z(t) = \sum_{k=1}^n \left[ a_k \cos(k \omega_0 t ) + b_k \sin(k \omega_0 t ) \right].
\label{eq_harm_timeser}\end{equation}
Learning this time series with a classical linear autoregressive model requires a memory capacity of $2n$ delays \citep{so_linear_2005,vaseghi_advanced_2008}.
Due to the nonlinearity generated by the encoding and the ancilla coupling of a QRC like \cref{eq_qrc_update}, this bound is only an upper limit in terms of actual number of required linear encoding features. Specifically, the number of ancilla qubits $N_A$ required to store the delays is much less than $2n$. While a detailed study will be presented elsewhere, we find here that $N_A=3$ and a well tuned memory capacity (via the TFIM parameters) suffices to learn $n\sim 10$ frequencies.
We use amplitude encoding [\cref{eq_enc_amp_sq}] and a resolution $\Delta t=0.2$ to feed the time samples $z_n=z(n\Delta t)$ into the QRC. The valid region of $\Delta t$ follows from the requirements that (i) the fading memory window covers a full period and (ii) the sampling is rate larger than $2 f\st{max}$ (Nyquist criterion).

\Cref{fig:qrc_opt_stateful}(a) illustrates the forecast horizons obtained from the optimized QRC using either the single optimal measurement operator $M^*$ (\cref{eq_Mopt}, dashed line) or decompositions into subsets of observables [\cref{eq_pauli_exp,eq_diag_spec}]. 
Using the full $M^*$ provides a stable forecast up to the maximum tested time, which is also reached when using all Pauli basis operators (dots).
When using the Pauli $Z$-observables, the maximum forecast time is only reached for certain random seeds used in constructing the time series (reflected by the rather large error bars).
The forecast can become unstable when using a small number of measurement operators, which essentially reflects numerical instabilities unrelated to the optimization process. 
These issues can, in principle, be alleviated by tuning the regularization parameter, input scale, or input noise (to be analyzed in detail elsewhere).

\paragraph*{Mackey-Glass time series.}
This model has nonlinear and non-Markovian dynamics and is defined by the delay differential equation \citep{mackey_oscillation_1977}
\beq \dot z(t) = \beta \frac{z(t-\theta)}{1 + z(t-\theta)^n} - \gamma z(t)
\label{eq_mcgl_DDE}\eeq 
with chaotic-regime parameters $\theta = 17$, $\beta = 0.2$, $\gamma = 0.1$, $n = 10$.
We evolve \cref{eq_mcgl_DDE} numerically with a time step $\Delta t=1$. 
Samples $z_n=z(n\Delta t)$ are scaled to the interval $[0,0.3]$ and injected with the amplitude encoding variant of \cref{eq_enc_amp_sqrt} into a QRC with TFIM-based unitary and $N_A=3$ ancilla qubits.
As seen in \cref{fig:qrc_opt_stateful}(b), the optimized QRC can forecast the Mackey-Glass time series up to $\geq 250$ time steps.
The Pauli-decomposition with slightly reduced observable count reaches even to $\sim 300$ time steps, possibly due to numerical stabilization effects not present when using the full operator $M^*$ or the Pauli-$Z$ decomposition.  
Note that the forecast horizon is necessarily finite due to the intrinsic Lyapunov instability of the Mackey-Glass system, with Lyapunov exponents reported of $\lambda_1 \simeq 6\times 10^{-3}$ \citep{sadath_approximating_2018}.

\section{Summary}
\label{sec_summary}

\begin{table}[tb]
    \centering
    \renewcommand{\arraystretch}{1.5}
    \begin{tabular}{p{0.18\textwidth} p{0.27\textwidth} p{0.5\textwidth}}
        \hline\hline
        \textbf{Step} & \textbf{Stateless (QELM)} & \textbf{Stateful (QRC)} \\
        \hline
        \textbf{1. Data} & Input pairs $\{(\xv_i, y_i)\}_{i=1}^P$ & Time series $\{\zv_t\}_{t=0}^{T}$; Targets $y_t = \zv_{t+1}$ \\
        & & Effective inputs $x_t := (\zv_{t-L}, \ldots, \zv_t)$\\
        & & $\{(x_t, y_t)\}_{t=L}^{T-1}$ can be treated as unordered samples  \\
        \hline
        \textbf{2. Encoding} & Map each $\xv_i$ to state: & To obtain $\rho(x_t)$, evolve for $\tau=t-L,\ldots,t$: \\
        & $\rho_i \equiv \rho(\xv_i)$ & $\eta(\tau) = U (\eta_I(\zv_\tau) \otimes \eta_M(\tau)) U^\dag$, $\eta_M(\tau+1)  = \tr_I \eta(\tau)$; \\
        &  & Then $\rho_t \equiv \rho(x_t) = \eta(t)$ \\
        \hline
        \textbf{3. Kernel} & \multicolumn{2}{c}{Gram matrix: $K_{ij} = K(x_i, x_j) = \tr(\rho_i \rho_j)$} \\
        \hline
        \textbf{4. Training} & \multicolumn{2}{c}{Solve for dual coefficients (least-squares): $\boldsymbol{\alpha}^* = (\bv{K} + \lambda \Imat)^{-1} \yv$} \\
        \hline
        \textbf{5. Optimal Observable} & \multicolumn{2}{c}{$M^* = \sum_{k=1}^P \alpha_k^* \rho_k$} \\
        \hline
        \textbf{6. Prediction} & For new input $\hat \xv$: & For new input $\hat x_t = (\hat \zv_{t-L}, \ldots, \hat \zv_t)$: \\
        & $f(\hat\xv) = \tr(M^* \rho(\hat \xv))$ & $f(\hat x_t) = \tr(M^* \rho(\hat x_t)) = \sum_{t'} \alpha_{t'}^* K(x_{t'}, \hat x_t)$ \\
        & $= \sum_i \alpha_i^* K(\xv_i, \hat\xv)$ &  \\
        \hline\hline
    \end{tabular}
    \caption{Algorithm flow for the kernel-based (dual) optimization of the measurement operator in stateless (memoryless) and stateful (recurrent) QRCs. Note that $\yv\in \reals^P$ is the vector of target samples and the expression for the dual coefficients $\alpha^*$ thus applies componentwise. In the stateful case, $P=T-L$ is the number of time steps available for training. For details, see \cref{app_qelm_kernel,app_qrc_stateful}.}
    \label{tab:algo_flow}
\end{table}

We have provided a framework, based on the connection between quantum kernel methods and supervised quantum ML models \citep{schuld_supervised_2021}, to optimize the measurement operator in quantum reservoir computers (QRCs).
This establishes a quantum kernel formulation on history space for recurrent QRCs, connecting the optimal readout observable to the kernel methods previously employed for memoryless quantum models.
Our method is thus applicable to both stateless (QELM) and stateful models, as summarized by the algorithm flow chart in \cref{tab:algo_flow}.
Since the essence of the method does not rely on the QRC-specific form \eqref{eq_QELM} of the output function, insights gained from this work can be transferred to other quantum ML models with a well-defined (and already trained) feature map $\xv\mapsto\rho(\xv)$, enhancing previous discussions of primal-dual approaches in QML \citep{huang_power_2021,jerbi_quantum_2023}.

Exact implementation of the optimal measurement operator $M^*$ on an $N$-qubit system requires measuring all $2^N$ $Z$-strings (or, equivalently, measuring in the computational basis) and a pre-measurement rotation of the quantum state by a global (training data-dependent) unitary. 
This unitary can be understood to capture the outcome otherwise achieved by training a parametrized QML model with fixed measurement operator.
Empirically, we find that using operator subsets (e.g., Pauli $Z$-strings) obtained from decomposing $M^*$ can sometimes improve forecasting performance compared to using the single operator $M^*$ directly. This could be related to the emergence of numerical instabilities or conditioning effects in certain QRC setups and should be investigated further.

Having determined an optimal measurement operator raises the question whether and to which extent the prediction performance of a QRC can be further improved.
As will be further discussed elsewhere, in QML models with a time-independent initial-state encoding strategy, the available features are fully determined by the encoding and are subsequently linearly processed by the quantum channels \footnote{For the product encoding schemes considered in \cref{sec_encodings}, the features are essentially polynomials of certain basis functions (e.g., Fourier exponentials or monomials in the input data).}.
This processing can manifest as information scrambling and reduce the prediction performance of QML models.
In the case of stateless QELMs, the unitary associated with the optimal observable can undo this unitary scrambling [see \cref{eq_diag_spec}], which is reflected by the invariance of the trained predictor \eqref{eq_QELM_kernel} under global conjugation.
Accordingly, for stateless QELMs with a unitary channel, optimization of the reservoir unitary can be traded against optimization of the measurement operator, up to the final measurement/readout parametrization.
By contrast, for stateful QRCs with decoherence/reset, or generally for non-unitary channels, optimizing the reservoir dynamics can change the kernel and thus the attainable predictor class.
Further meaningful optimizations pertain to the encoding, memory depth and possibly hand-engineered readout features.
Alternatively to determining the optimal measurement operator, one may use temporal multiplexing \citep{fujii_harnessing_2017,fujii_quantum_2020} or measuring over a complete measurement basis (full state tomography).

We have numerically demonstrated the feasibility of the optimization for QRCs with $N\lesssim 10$ qubits, noting that, for our considered datasets, it suffices to use a representative subset of the full training samples to evaluate the kernel matrix.
Application to larger systems is instead bottlenecked by memory consumption of the $4^N$-dimensional operator matrix and requires further methodological developments.
Another avenue for future work is a closer look into generalization capability entailed by the optimal measurement operator, since both primal and dual optimization methods provide only error bounds on the training set \citep{canatar_bandwidth_2022,gross_expressivity_2025,li_enhancing_2025}.


\appendix 

\section{QELM as a kernel method}
\label{app_qelm_kernel}

In the following, we describe in detail the mapping between a QRC/QELM and a quantum kernel method and the ensuing measurement operator optimization.
This builds on standard material of quantum ML \citep{shawe-taylor_kernel_2004,nielsen_quantum_2010,schuld_supervised_2021,schuld_machine_2021,kempkes_double_2026}, adapted specifically to the case of QRCs. 
In the following, we focus on the QELM formulation and discuss the stateful QRC separately in \cref{app_qrc_stateful} below.

We consider an $N$-qubit system with Hilbert space dimension $D=2^N$, on which a density matrix acts as a feature map $\xv \mapsto \rho(\xv)$. On the real vector space of Hermitian operators $\Herm(\Hcal)$ we use the Hilbert--Schmidt inner product $\bra A, B \ket = \tr (A B)$ with induced norm $(|A|\ut{HS})^2=\tr (A^2)$. A QELM with a linear readout layer specifies an observable $M\in \Herm(\Hcal)$ and predicts the output
\beq
f_M(\xv) = \tr \big(M \rho(\xv)\big).
\label{eq_qelm-basic}
\eeq
Given training data ${(\xv_i,y_i)}_{i=1}^P$, we optimize the measurement operator $M$ using a Tikhonov-regularized loss function $\Lcal$
\beq
\min_{M\in\Herm(\Hcal)} \sum_{i=1}^P \Lcal \big(f_M(\xv_i),y_i\big) + \frac{\lambda}{2}(|M|\ut{HS})^2,
\qquad \lambda>0.
\label{eq_qelm-erm}
\eeq
Typical loss functions are the squared-error loss $\Lcal(\hat y,y) = \frac{1}{2}(\hat y-y)^2$ (regression problem) and the hinge loss $\Lcal(\hat y,y) = \max(0,1-y\hat y)$ (classification problem).

The representer property holds directly in this operator space \citep{schuld_supervised_2021,kubler_inductive_2021}. Decompose any $M$ as $M = M_\parallel + M_\perp$, where $M_\parallel \in \mathrm{span}\{\rho(\xv_1),\ldots,\rho(\xv_P)\}$ and $M_\perp$ is Hilbert--Schmidt orthogonal to this span. Then $\tr (M_\perp \rho(\xv_i))=0$ for all $i$, so the data-fit term in \cref{eq_qelm-erm} depends only on $M_\parallel$, while $(|M|\ut{HS})^2 = (|M_\parallel|\ut{HS})^2 + (|M_\perp|\ut{HS})^2$. Replacing $M$ by $M_\parallel$ strictly decreases the objective unless $M_\perp=0$. Therefore, there exists an optimal observable $M^*$ in the span of the training data, i.e.,
\beq
M^* = \sum_{i=1}^P \alpha_i^* \rho(\xv_i).
\label{eq_rep-thm}
\eeq

Substituting \cref{eq_rep-thm} into \cref{eq_qelm-erm} produces an optimization over the coefficients $\boldsymbol{\alpha}=(\alpha_1,\ldots,\alpha_P)^\top$. Define the Gram matrix $\bv{K}\in\reals^{P\times P}$ with $K_{ij} = \tr (\rho(\xv_i)\rho(\xv_j))$ and the label vector $\yv=(y_1,\ldots,y_P)^\top$. For each training point,
\beq
f_{M}(\xv_i) = \tr \Big(\sum_{j=1}^P \alpha_j \rho(\xv_j)\, \rho(\xv_i)\Big) = \sum_{j=1}^P \alpha_j K_{ij},
\label{eq_f-on-train}
\eeq
and the Hilbert--Schmidt norm evaluates to
\beq
(|M|\ut{HS})^2 = \tr \Big(\sum_{i} \alpha_i \rho(\xv_i)\, \sum_{j} \alpha_j \rho(\xv_j)\Big) = \boldsymbol{\alpha}^\top \bv{K} \boldsymbol{\alpha}.
\label{eq_hs-alpha}
\eeq
Hence, \cref{eq_qelm-erm} reduces to the convex optimization problem
\beq
\min_{\boldsymbol{\alpha}\in\reals^P} \sum_{i=1}^P \Lcal\big([\mathbf{K}\boldsymbol{\alpha}]_i, y_i\big) + \frac{\lambda}{2} \boldsymbol{\alpha}^\top \mathbf{K} \boldsymbol{\alpha} .
\label{eq_krr-objective}
\eeq
For the squared-error loss, this is the kernel ridge regression objective
\beq
\min_{\boldsymbol{\alpha}\in\reals^P} \frac{1}{2}\,\|\bv{K}\boldsymbol{\alpha} - \yv\|_2^2 + \frac{\lambda}{2} \boldsymbol{\alpha}^\top \bv{K} \boldsymbol{\alpha},
\label{eq_krr-objective-mse}
\eeq
which has a closed solution $\boldsymbol{\alpha}^*$ determined by the linear system $(\bv{K} + \lambda \Imat)\, \boldsymbol{\alpha}^* = \yv$.
For the hinge loss, this becomes equivalent to the dual ``quadratic programming problem'' \citep{bishop_pattern_2006}:
\beq
\max_{\boldsymbol{\beta}\in\reals^P} \sum_{i=1}^P \beta_i - \frac{1}{2\lambda} \sum_{i,j=1}^P \beta_i \beta_j y_i y_j K_{ij}
\quad \text{subject to} \quad 0 \le \beta_i \le 1,
\label{eq_svm_dual}
\eeq
with $\alpha_i^* = y_i \beta_i^*/\lambda$ (and no bias).
In this case, the solution coefficients $\boldsymbol{\alpha}^*$ are not available in closed form, but can be computed numerically.

For any loss function $\Lcal$ in \cref{eq_qelm-erm}, the optimal observable and predictor admit the kernel representation
\beq
M^* = \sum_{i=1}^P \alpha_i^*\, \rho(\xv_i),
\qquad
f^*(\xv) = \tr \big(M^* \rho(\xv)\big) = \sum_{i=1}^P \alpha_i^*\, K(\xv,\xv_i),
\label{eq_final-equivalence}
\eeq
with kernel
\beq
K(\xv,\xv') = \tr \big(\rho(\xv)\rho(\xv')\big).
\label{eq_hs-kernel}
\eeq
Thus, for the squared-error loss, measurement optimization for a QELM is exactly kernel ridge regression with kernel \cref{eq_hs-kernel}, and the QELM predictor equals the kernel predictor on all inputs when the optimal observable \cref{eq_rep-thm} is used.

Expanding $M^*$ in terms of an orthogonal operator basis $\{M_k\}$ with $G_{kl}\equiv \tr(M_k M_l)=g_k \delta_{kl}$, yields $M^*=\sum_k w_k M_k$ with 
\beq w_k^* = \frac{1}{g_k} \tr(M_k M^*) = \frac{1}{g_k} \sum_{i=1}^P \alpha_i^* \tr(M_k \rho(\xv_i)).
\label{eq_primal_dual_weights}\eeq
For the Pauli basis, $g_k = 2^N$. 
Thus, taking 
\beq \phi_k(\xv)=\tr(M_k \rho(\xv))
\eeq 
as the feature map \footnote{For a complete orthonormal operator basis, $\phi_k$ represents the expansion coefficients of the feature map $\rho(\xv)$ under the HS-inner product $\tr(\cdot)$ and thus becomes identical to the \emph{constrained} feature map $\varphi$ defined in \cref{eq_primal_weights,tab:qrc_opt_forms}.} and $\Phi_{ik} = \phi_k(\xv_i)$ as the feature matrix $\Phi\in\reals^{P\times m}$ (with rows $\phi(\xv_j)^T$), the QELM predictor can be written in the form of a classical ELM [see \cref{eq_elm,eq_elm_wts_sol_dual} below]:
\beq f^*(\xv) = \phi(\xv)^T \wv^* ,\qquad \wv^* =  G^{-1} \Phi^T \boldsymbol{\alpha}^*.
\label{eq_qelm_feature_map}\eeq
Using the HS completeness identity $\sum_{k,l} \tr(M_kA) (G^{-1})_{kl}\tr(M_l B) = \tr(A B)$ for any Hermitian operators $A,B$, it is easy to see that, indeed, $f^*(\xv)= \sum_i \alpha_i^* K(\xv,\xv_i)$.

Consider now the case where $\{M_k\}_{k=1}^\numM$ is \emph{not a complete basis}, i.e., one cannot generally assume that there exists a $\wv$ with $\sum_k w_k M_k = M^*$. The natural generalization of \cref{eq_primal_dual_weights} is to find a $\wv$ that minimizes the (squared) Hilbert--Schmidt distance $\|\sum_k w_k M_k - M^*\|_\text{HS}^2 = \tr \big[(\sum_k w_k M_k - M^*)^2\big]$. Solving this optimization problem yields $\wv = G^+ \Phi^T \alpha^*$, or component-wise
\beq  w_k = \sum_{l=1}^\numM (G^+)_{kl} \sum_{j=1}^P \alpha_j^* \tr(M_l \rho(\xv_j)).
\label{eq_qelm_primwts_incomp}\eeq

\section{Generalization to multiple outputs/classes}
\label{app_multioutput}

Extension of the framework in \cref{app_qelm_kernel} to multiple outputs/classes is straightforward. 
Consider first \emph{regression} with squared-error loss: for $C$-dimensional targets $\yv_i\in\reals^C$, one assembles $\Yv=[\yv_1,\ldots,\yv_P]^\top\in\reals^{P\times C}$, solves $\Av^*=(\bv{K}+\lambda \Imat)^{-1}\Yv$, sets $M_c^*=\sum_{i=1}^P \alpha_{ic}^* \rho(\xv_i)$, and predicts $f_c^*(\xv)=\sum_{i=1}^P \alpha_{ic}^* K(\xv,\xv_i)$. 

For multi-class \emph{classification} with labels $y_i\in\{1,\ldots,C\}$, one can use one-hot targets and interpret the components $f_c^*(\xv)$ as class scores, predicting $\hat y(\xv)=\arg\max_{c} f_c^*(\xv)$ (or applying a softmax to obtain class scores). 
One thus replaces the single observable $M$ by $C$ observables $\{M_c\}_{c=1}^C$ (equivalently a coefficient matrix $\Av\in\reals^{P\times C}$) with regularizer $\frac{\lambda}{2}\sum_c (|M_c|\ut{HS})^2$. The representer form then holds component-wise as $M_c^*=\sum_i \alpha_{ic}^*\rho(\xv_i)$, and one-vs-rest hinge/logistic training reduces to $C$ standard kernel problems with the same Gram matrix $\bv{K}$.

\section{Diagonalization}
\label{sec_diag_remarks}
Because the computational basis is the joint eigenbasis of all single-qubit $Z$ operators, any diagonal operator admits an expansion in commuting Pauli-Z strings:
\beq
\Lambda = \sum_{\sv\in\{0,1\}^N} \beta_\sv\, Z_\sv,\qquad Z_\sv := \bigotimes_{j=1}^N (\sigma_z)^{s_j},
\label{eq_Z_expansion}
\eeq
with coefficients given by a Walsh–Hadamard transform of the spectrum $\{\lambda_b\}$ (see, e.g., \cite{welch_efficient_2014,georges_pauli_2025}),
\beq
\beta_\sv = 2^{-N} \tr (\Lambda Z_\sv) = 2^{-N} \sum_{\bv{b}\in\{0,1\}^N} \lambda_\bv{b}\, (-1)^{ \sv \cdot \bv{b}}.
\label{eq_walsh_hadamard}
\eeq
Consequently, the optimal measurement operator $M^*$ can be expanded in terms of $Z$-basis measurement operators $Z_\sv$:
\beq
M^* = V \Lambda V^\dag = \sum_{\sv\in\{0,1\}^N} \beta_\sv\, V Z_\sv V^\dag.
\label{eq_pullback_Z}
\eeq
After the pre-rotation $V^\dag$, all $Z_\sv$ can be evaluated from the same computational-basis samples by taking products of single-qubit outcomes; combining these estimates with weights $\{\beta_\sv\}$ yields the same predictor as the eigenvalue-weighted average in \cref{eq_diag_expect}. Both procedures use a single circuit per shot.

\section{Pauli basis truncation}
\label{app_truncation_error}

Assume we measure only a subset of Pauli observables in the expansion of $M^*$ [\cref{eq_pauli_exp}]. 
Let the corresponding index set be $S\subset \{0,\ldots,D^2-1\}$ and define the truncated observable $M\st{trunc}=\sum_{k\in S} c_k P_k$. 
By \cref{eq_qelm-basic}, the pointwise prediction error on any input $\xv$ is
$\Delta f(\xv) = \tr \big((M^*-M\st{trunc}) \rho(\xv)\big)$.
Using the Cauchy–Schwarz inequality in the Hilbert--Schmidt inner product yields $|\Delta f(\xv)| \le |M^*-M\st{trunc}|\ut{HS}\, |\rho(\xv)|\ut{HS}$.
Since for any feature state $|\rho(\xv)|\ut{HS} \le 1$, the worst-case prediction error is bounded by the magnitude of the omitted coefficients:
\beq
|\Delta f(\xv)|\leq \|M^*-M\st{trunc}\|\st{HS} = \Big(2^N \sum_{k\notin S} c_k^2\Big)^{1/2},
\label{eq_tail}
\eeq
where we used orthogonality of the basis.

\section{Classical ELM}
\label{sec_classical_elm}

It is instructive to review the classical extreme learning machine (ELM) \citep{huang_extreme_2006} as a paradigmatic example for a kernel formulation.
The ELM is essentially a generalized linear regression model \citep{bishop_pattern_2006,murphy_probabilistic_2022} with a set of $m$ fixed features $\phi(\xv)\in\reals^m$ and trainable weights $\wv\in \reals^m$:
\beq f(\xv) = \phi(\xv)^T \wv.
\label{eq_elm}\eeq
Recall that a QELM can be mapped to this form by identifying $\phi_k(\xv) = \tr[M_k \rho(\xv)]$ for a set of $m$ observables $\{M_k\}$ [see \cref{eq_qelm_feature_map}].
Assuming a least-squares loss,
\beq \Lcal(\{\xv_j, y_j\})= \onehalf \sum_{j=1}^P \left[y_j - f(\xv_j)\right]^2 + \frac{\lambda}{2} \| \wv \|^2,
\label{eq_elm_loss}
\eeq 
and $P$ samples, the optimal solution follows as
\beq \wv^* = (\Phi^T \Phi + \lambda \Imat)^{-1} \Phi^T \yv
\label{eq_elm_wts_sol}\eeq 
with the feature matrix $\Phi\in\reals^{P\times m}$ (with rows $\phi(\xv_j)^T$) and sample vector $\yv$.
From the feature vector $\phi(\xv)$, an associated kernel can be defined as \citep{bishop_pattern_2006,hofmann_kernel_2008,murphy_probabilistic_2022}
\beq k(\xv,\xv') = \phi(\xv)^T \phi(\xv') \in \reals,
\eeq 
which gives rise to the Gram matrix $\bv{K}\in\reals^{P\times P}$ with $K_{ij}=k(\xv_i,\xv_j) = (\Phi \Phi^T)_{ij}$.
Noting that $(\Phi^T\Phi+\lambda\Imat)^{-1}\Phi^T = \Phi^T (\Phi\Phi^T+\lambda\Imat)^{-1}$, the primal solution in \cref{eq_elm_wts_sol} can be rewritten as
\beq 
\wv^* = \Phi^T \boldsymbol \alpha^* = \sum_{j=1}^P \alpha_j^* \, \phi(\xv_j)
\label{eq_elm_wts_sol_dual}\eeq 
with the dual coefficients
\beq \boldsymbol{\alpha}^* = (\bv{K}+\lambda \Imat)^{-1}\yv.
\eeq
Accordingly, the trained ELM predictor [\cref{eq_elm}] can be equivalently written as a kernel model:
\beq f(\xv) = \phi(\xv)^T \wv^* = \phi(\xv)^T\Phi^T \boldsymbol{\alpha}^* = \sum_{j=1}^P \phi(\xv)^T \phi(\xv_j) \alpha_j = \sum_{j=1}^P \alpha_j k(\xv,\xv_j).
\eeq 
The mapping \eqref{eq_elm_wts_sol_dual} shows that, even in a potentially infinite feature space, the representer theorem \citep{shawe-taylor_kernel_2004,hofmann_kernel_2008} guarantees that the optimal solution can be expressed in terms of a finite number of samples.
Primal and dual optimizations thus render the same set of functions.
Note that the expressions \eqref{eq_elm_wts_sol} and \eqref{eq_elm_wts_sol_dual} pertain to the least-squares loss \eqref{eq_elm_loss}. For other loss functions, the solutions typically have to be determined numerically.

\section{QRC with stateful reservoir}
\label{app_qrc_stateful}

We describe here in some more detail how the kernel-based optimization can be applied to the case of a QRC with internal memory, complementing \cref{app_qelm_kernel} and demonstrating the viability of prescription \cref{eq_eff_seq_input}. 
Let the input register of the QRC have $N_I$ qubits and the memory register $N_A$ qubits.  
At time $t$, a classical input $\zv_t \in \mathbb{R}^{d'}$ (typically $d'=N_I$) is encoded via the density matrix $\eta_I(\zv_t)$. The joint state $\eta_I(\zv_t) \otimes \eta_M(t)$ is time-evolved by a unitary $U$, giving 
\beq \eta(t)= U \left(\eta_I(\zv_t) \otimes \eta_M(t)\right) U^\dagger.
\label{eq_qrc_state_evo}\eeq 
Subsequently, the input is measured/reset and discarded, and the memory alone is carried to the next step, described by the CPTP map
\beq
\eta_M(t+1)  = \tr_I\!\left[ \eta(t)\right].
\label{eq_qrc_state_mem}\eeq

The stateful QRC dynamics above induces the quantum feature map
\beq
(\zv_0,\dots,\zv_t)=:x_t \longmapsto \rho(x_t)=\eta(t)\in\Herm(\Hcal),
\label{eq_full_eff_input}\eeq
where $\rho(x_t)$ is obtained after sequentially processing the $\{\zv_{t'}\}$ via \cref{eq_qrc_state_evo,eq_qrc_state_mem}.
Due to the fading memory of the reservoir, the dependence on inputs earlier than time $t-L$ is negligible. We assume that any transient associated with the reservoir initialization has already decayed before the observed sequence begins, e.g. by driving the reservoir for times $t<0$. We therefore work, instead of \eqref{eq_full_eff_input}, with the effective input history (window)
\beq
x_t := (\mathbf{z}_{t-L},\ldots,\mathbf{z}_t)\in (\mathbb{R}^{d'})^{L+1},\qquad t=L,\ldots,T-1,
\eeq
and define $P=T-L$ training samples.
Let $\rho(x_t)$ denote the reservoir state obtained from driving the reservoir with that history from some fixed initialization (the precise initialization is irrelevant by assumption).
The corresponding readout is
\beq
f_M(x)=\tr\bigl(M\rho(x)\bigr).
\eeq
This exactly corresponds to \cref{eq_QELM}, except that the state $\rho(x_t)$ is to be determined by evolving the reservoir over the input sequence $(\zv_{t-L},\ldots,\zv_t)$.

For targets $\{y_t\}_{t=L}^{T-1}$, we consider the optimization problem
\beq
\min_{M\in\Herm(\Hcal)}
\frac{1}{2}\sum_{t=L}^{T-1}\bigl(\tr(M\rho(x_t)) - y_t\bigr)^2
+ \frac{\lambda}{2}\tr(M^2),
\eeq
which can be straightforwardly extended to other loss functions.
By the representer theorem in $(\Herm(\Hcal),\langle\cdot,\cdot\rangle_{\mathrm{HS}})$, there exists an optimal observable \citep{shawe-taylor_kernel_2004,hofmann_kernel_2008},
\beq
M^*=\sum_{t=L}^{T-1}\alpha_t^* \rho(x_t),
\eeq
which is the stateful analogue of \cref{eq_Mopt}.
We now define the kernel on histories/windows
\beq
K(x,x'):=\tr\bigl(\rho(x)\rho(x')\bigr),
\eeq
and the Gram matrix $K_{tt'}:=K(x_t,x_{t'})$. Then the coefficients satisfy
\beq
\boldsymbol{\alpha}^*=(\mathbf{K}+\lambda\mathbb{I})^{-1}\mathbf{y},
\eeq
with $\boldsymbol{\alpha}^* = (\alpha_t^*)_{t=L}^{T-1}$ and the target vector $\mathbf{y} = (y_t)_{t=L}^{T-1}\in\mathbb{R}^P$.
For any test history/window $x$ (obtained from any test sequence driven through the same reservoir),
\beq
f^*(x)=\sum_{t=L}^{T-1}\alpha_t^* K(x_t,x).
\eeq
For a fixed driving sequence, one may write $K(t,t')$ as shorthand for $K(x_t,x_{t'})$.
This is the direct stateful counterpart of the kernel representation in \cref{eq_QELM_kernel}, with the feature map $\rho(x)$ replaced by the stateful reservoir map $(\zv_{t-L},\dots,\zv_t)=:x_t \mapsto\rho(x_t)$. For vector-valued targets $\yv_t\in\mathbb{R}^C$ (for example, when predicting all components of $\zv_{t+1}$), one can apply the same construction component-wise, as in the stateless case discussed below \cref{eq_opt_kernel_params}.

\section{Product state encodings and measurements}
\label{app_prodstates}

If $\rho=\bigotimes_{j=1}^N \rho_j$ and $P=\bigotimes_{j=1}^N \sigma^{p_j}$ is a Pauli string, then the measurement on the $N$-qubit product state gives
\beq\tr(P\rho)=\prod_{j=1}^N \tr(\sigma^{p_j}\rho_j).
\label{eq_prodstate}\eeq
Any single-qubit state can be expanded in the Pauli basis as
\beq
\rho_j = \onehalf\Bigl(\Imat + \sum_{p\in\{x,y,z\}} \tr(\sigma^p\rho_j)\,\sigma^p\Bigr).
\label{eq_pauli_exp_singlequbit}\eeq
For single-qubit rotational encoding \eqref{eq_enc_rot_Y} one has
\beq
\rho_j(z)=|\psi_{\text{rot}}(z)\ket\bra\psi_{\text{rot}}(z)|
=
\begin{pmatrix}
\cos^2(\pi z) & \cos(\pi z)\sin(\pi z)\\
\cos(\pi z)\sin(\pi z) & \sin^2(\pi z)
\end{pmatrix}
= \onehalf\Bigl(\Imat + \sin(2\pi z)\,\sigma^x + \cos(2\pi z)\,\sigma^z\Bigr),
\label{eq_rho_rot_Y}\eeq
while for amplitude encoding \eqref{eq_enc_amp_sqrt}:
\beq
\rho_j(z)=
\begin{pmatrix}
z & \sqrt{z(1-z)}\\
\sqrt{z(1-z)} & 1-z
\end{pmatrix}
= \onehalf\Bigl(\Imat + 2\sqrt{z(1-z)}\,\sigma^x + (2z-1)\,\sigma^z\Bigr).
\label{eq_rho_amplsqrt}\eeq
\Cref{tab_prodstate_expect} summarizes the single-qubit expectation values obtained for rotational and amplitude encoding.
\begin{table}[ht]
\centering
\begin{tabular}{@{}c|c|c@{}}
\hline
 & \multicolumn{2}{c}{$\tr(\sigma^p\rho_j)$}\\
$\sigma^p$ & rotational & amplitude\\
\hline
$\mathbb{I}$ & $1$ & $1$\\
$\sigma^x$ & $\sin(2\pi z)$ & $2\sqrt{z(1-z)}$\\
$\sigma^y$ & $0$ & $0$\\
$\sigma^z$ & $\cos(2\pi z)$ & $2z-1$\\
\hline
\end{tabular}
\caption{Single-qubit expectation values for rotational and amplitude encoding.}
\label{tab_prodstate_expect}
\end{table}

This implies, in particular, that measuring the above encoded product states with a Pauli string $P$ containing a $Y$-factor will always yield zero \footnote{This does not apply to the dense encoding in \cref{eq_enc_rot_dense}.}.
Action of a unitary breaks \cref{eq_prodstate} as it typically spreads any $\sigma^p$ into a mixture of all other Paulis, such that measuring a $Y$ then will result in nonzero contributions.


\bibliography{bibliography}

\end{document}